\documentclass[aps,prd,10pt,nofootinbib,superscriptaddress,altaffilletter,
twocolumn,floatfix]{revtex4-2}

\DeclareUnicodeCharacter{2212}{\ensuremath{-}}
\usepackage{amsmath,amssymb}
\usepackage{graphicx}
\usepackage{multirow,booktabs}
\usepackage{hyperref}
\usepackage{orcidlink}
\usepackage{xcolor}

\usepackage[]{latexsym}
\usepackage[english]{babel}
\usepackage{graphicx}
\usepackage{amsmath}
\hypersetup{
  colorlinks=true,
  linkcolor=blue,
  citecolor=blue,
  urlcolor=blue,
  filecolor=blue,
  linktoc=all
}
\allowdisplaybreaks

\begin{document}

\title{Cosmological Constraints on Interacting Scalar-Field Dark Matter and Early Dark Energy mediated by a Kinetic Kernel}

\author{Gabriela Garcia-Arroyo\,\orcidlink{0000-0002-0599-7036}}
\email{arroyo@icf.unam.mx}\affiliation{Instituto de Ciencias Físicas,
Universidad Nacional Autónoma de México,
62210, Cuernavaca, Morelos, México.}

\author{L. Arturo Ureña-López\,\orcidlink{0000-0001-9752-2830}}
\email{lurena@ugto.mx}\affiliation{Departamento de Física, DCI, Campus León, 
Universidad de Guanajuato, 37150, León, Guanajuato, México.}

\author{J. Alberto V\'azquez\,\orcidlink{0000-0002-7401-0864}}
\email{javazquez@icf.unam.mx}\affiliation{Instituto de Ciencias Físicas,
Universidad Nacional Autónoma de México,
62210, Cuernavaca, Morelos, México.}


\begin{abstract}
In this work, we build a general framework for constraining interacting scalar fields with cosmological observations, applicable to a broad family of scalar--field potentials and interaction kernels. As a representative example, we derive constraints on a model in which  dark matter is described by a scalar field $\phi$ with a quadratic potential and dark energy by a scalar field $\psi$ with the Albrecht--Skordis potential, considering both the uncoupled case and a more general scenario in which the exchange of energy occurs through an interaction kernel proportional to the kinetic energies of the scalar fields $Q \propto \dot{\phi}^2 \dot{\psi}^2$. The scalar field parameters are constrained by using \textit{Planck} 2018 CMB data, DESI BAO measurements, Ly-$\alpha$ forest constraints, and SNeIa observations, and we find that the scalar field parameters are driven toward their $\Lambda$CDM-like limits; the early dark energy fraction is largely suppressed, and the SFDM mass is bounded to values for which the field behaves close to cold dark matter. In the interacting case, the results show no significant preference for the transfer of energy between dark energy and dark matter at 95\% C.L. In general, the model keeps $H_0$ close to the Planck-$\Lambda$CDM range and provides best-fit $\chi^2$ values comparable to those of the corresponding $\Lambda$CDM fits.
\end{abstract}

\maketitle

\section{Introduction}

The standard cosmological model, known as $\Lambda$CDM, has proven to be remarkably successful in describing a wide range of cosmological observations across different epochs and distance scales \cite{SupernovaSearchTeam:1998fmf, SupernovaCosmologyProject:1998vns, BOSS:2012bus, Hildebrandt:2016iqg, Planck:2019nip}. However, despite its success, one of its main challenges remains the lack of understanding of the fundamental nature of the dark sector. 
To address this limitation, several theoretical alternatives have been proposed, seeking to offer a more natural description of dark matter (DM) and dark energy (DE). Some of the most compelling possibilities are models based on scalar fields, where one or both parts of the dark sector are represented by dynamical scalar fields \cite{Ratra:1987rm, Caldwell:1997ii, Matos:2025pkk}. In these scenarios, the fields evolve according to the Klein–Gordon equation under the influence of a potential, enabling richer dynamics than those allowed by a cosmological constant or a pressureless fluid.
Furthermore, it is common to allow energy or momentum exchange between DE and DM, especially when both components are treated as dynamic fluids \cite{Valiviita:2008iv, Bolotin:2013jpa,Wang:2016lxa,DiValentino:2019ffd, Carrion:2024itc, Giare:2024smz,Wang:2024vmw, Figueruelo:2026eis}.
This transfer is typically encoded in a phenomenological source term $Q$ that appears with opposite signs in the conservation equations of the two dark components, preserving the total dark-sector energy budget while redistributing energy between them \cite{Adil:2026ift}.
Within scalar field frameworks,  the usual approach is to describe only one dark component as a scalar field, while the other remains as a perfect fluid, despite the fact that both sectors may be theoretically motivated by scalar degrees of freedom \cite{Pourtsidou:2013nha, Costa:2014pba, Boehmer:2015kta, Boehmer:2015sha, An:2018vzw, Perez:2021cvg, Gomez-Valent:2022bku,vandeBruck:2022xbk, Roy:2023uhc, Aboubrahim:2024spa, Kritpetch:2024rgi,Li:2026xaz}.

Throughout this work, we build a framework in which dark matter and dark energy are described by scalar fields that are able to interact via energy transfer. Based on our previous analysis \cite{Garcia-Arroyo:2024tqq}, 
the central idea is to substitute the standard cold dark matter component with an ultralight scalar field (SFDM) with an associated mass $m_{\phi}\sim 10^{-22} \rm{eV} $ \cite{Lee:1995af, Matos:1998vk, Matos:1999et, Hu:2000ke, Urena-Lopez:2023ngt}. 
Such field can effectively behave as a pressureless matter on large scales while suppressing small-scale structure formation below a characteristic Jeans scale \cite{Urena-Lopez:2015gur}.
This kind of behavior may be achieved by different choices of the associated potential \cite{Matos2009, Matos2009DynamicsOS, Suarez:2013iw, Ross:2016hyb, Cedeno:2017sou, Urena-Lopez:2019xri}, but as a proof of concept, we restrict ourselves to the simplest quadratic potential, which captures the leading oscillatory dynamics and provides a clean baseline for isolating the impact of the coupling to the dark-energy field.
On the other hand, as a viable alternative to the cosmological constant, the corresponding scalar field is generally referred to as quintessence \cite{Vazquez:2020ani, Steinhardt:2003st, Linder:2007wa, Tsujikawa:2013fta, PhysRevD.98.063530, Ibitoye:2026whe}, but depending on the type of potential, it may also behave as a phantom field \cite{Vazquez:2023kyx, Caldwell:1999ew,  Caldwell:2003vq, Copeland:2006wr}. From an extended variety of potentials, we select the Albrecht-Skordis (AS) potential \cite{Skordis:2000dz, Barrow:2000nc}, which behaves like quintessence.
A key property of the AS is that it belongs to the class of scaling potentials, allowing the field to contribute with a small, but non-negligible, fraction of the energy density in the early universe,  making it a natural candidate for describing early dark energy (EDE). As the field eventually settles near the minimum of the potential, it can also mimic a cosmological constant at later times.
This feature is particularly advantageous since, in many EDE scenarios based on axion-like or generalized axion-like \cite{Bouhmadi-Lopez:2025wxo} potentials, it is necessary to introduce a cosmological constant to reproduce the late-time accelerated expansion of the universe \cite{Poulin:2018dzj, Poulin:2018cxd,Kamionkowski:2022pkx, McDonough:2023qcu}.
Moreover, in addition to their gravitational interaction, the fields can also interact directly through an energy–transfer term~$Q$, allowing for a bidirectional exchange  between them.
This framework is implemented in a modified version of the publicly available Boltzmann solver \texttt{CLASS} that provides a way to test whether the interplay between SFDM and EDE parameters under the influence of a direct coupling can leave an imprint on current cosmological data. A similar two-field approach for the dark sector was recently explored in \cite{Rahimy:2025iyj}, where the authors studied coupled dynamics and classified the resulting cosmological behaviors.
\\

Previously, on \cite{Garcia-Arroyo:2024tqq}, we focused on theoretical predictions for background evolution and linear perturbations, including the cosmic microwave background (CMB) temperature and matter power spectra (MPS). We found that the presence of a SFDM component, especially when modeled with a trigonometric potential, can partially offset the suppression in the matter power spectrum typically induced by EDE. 
However, in the present work, we extend that analysis by performing a full Bayesian parameter estimation, incorporating recent cosmological datasets. Specifically, we used \textit{Planck} 2018 CMB data~\cite{Planck:2019nip}, Baryon Acoustic Oscillations (BAO) measurements from DESI~\cite{DESI:2025zgx}, the matter power spectrum inferred from the Ly-$\alpha$ forest of eBOSS DR14~\cite{Chabanier:2019eai}, and SNeIa data from Pantheon+SH0ES~\cite{Scolnic:2021amr, Brout:2022vxf}. The addition of Ly-$\alpha$ data is crucial for constraining the SFDM mass, as CMB data alone do not provide sufficient small-scale sensitivity to break degeneracies with other cosmological parameters~\cite{Irsic:2017yje,Armengaud:2017nkf}. By combining these complementary datasets, we aim to place tighter constraints on the parameters describing the scalar field potentials, the strength of the interaction, and the allowed mass range for the SFDM component, thereby assessing the viability of this framework in addressing current cosmological tensions.
\\

The paper is organized as follows. In Sec.~\ref{sec:theory}, we present the theoretical description of the cosmological models. In Sect.~\ref{sec:methodology}, we detail the methodology and datasets used throughout this work. The main results of the Bayesian analysis are shown in Sec.~\ref{sec:results}, and in Sec.~\ref{sec:conclusions}, we summarize our conclusions and discuss the implications of our findings.

\section{Theoretical framework}\label{sec:theory}

The dark sector is described by two canonical scalar fields, $\phi$ for dark matter and $\psi$ for dark energy, evolving in a flat FLRW metric with an additional interaction kernel, $Q$. Their background dynamics follow the Klein-Gordon (KG) equations~\cite{Garcia-Arroyo:2024tqq}:
\begin{align}
\ddot{\phi} + 3H\dot{\phi} + \partial_\phi V(\phi) &= \frac{Q}{\dot{\phi}}, \\
\ddot{\psi} + 3H\dot{\psi} + \partial_\psi V(\psi) &= -\frac{Q}{\dot{\psi}} ,
\end{align}
where overdots denote derivatives with respect to cosmic time, $H$ is the Hubble expansion rate, $V(\phi)$ and $V(\psi)$ are the scalar-field potentials for each component, $\partial_x V \equiv \partial V/\partial x$ represents the derivative of the potential with respect to each field $x$, and $Q$ is the interaction kernel that parametrizes direct energy exchange between the two fields.
\\

At the perturbative level, the scalar fluctuations in Fourier space, and synchronous gauge, satisfy the linearized KG equations:
\begin{align}
\ddot{\delta\phi}+3H\dot{\delta\phi}+\big(k^2+\partial^2_\phi V\big)\delta\phi+\tfrac12 \dot{h}\,\dot{\phi}
&=\delta\!\left(\frac{Q}{\dot{\phi}}\right), \\
\ddot{\delta\psi}+3H\dot{\delta\psi}+\big(k^2+\partial^2_\psi V\big)\delta\psi+\tfrac12 \dot{h}\,\dot{\psi}
&=-\delta\!\left(\frac{Q}{\dot{\psi}}\right),
\end{align}
where $k$ is the comoving wavenumber, $\delta\phi$ and $\delta\psi$ are the field perturbations, and $h$ is the synchronous metric perturbation trace. Here, $\partial^2 V$ denotes the second derivatives of the corresponding potentials, and the interaction enters through the perturbed source terms $\delta(Q/\dot{\phi})$ and $\delta(Q/\dot{\psi})$, respectively.  
\\

\noindent
\textbf{Dark Matter contribution.}\\
\noindent The scalar field $\phi$, associated with dark matter, is assumed to evolve under a quadratic potential:
\begin{equation}\label{eq:qua_potential}
V(\phi) = \tfrac{1}{2} m_\phi^2 \phi^2 ,
\end{equation}
where $m_\phi$ is the associated scalar mass. This equation captures the effective dynamics of a wide class of SFDM models once the field enters the dark matter regime, in which rapid oscillations around the minimum lead to an average pressureless behavior, enabling the scalar field to mimic cold dark matter on large scales while introducing a suppression of small-scale structure in the matter power spectrum~\cite{Urena-Lopez:2015gur}.
\\

\noindent
\textbf{Dark Energy contribution.}\\
\noindent In contrast, dark energy is described by a quintessence field $\psi$, evolving through the Albrecht--Skordis potential:
\begin{equation}\label{eq:AS_potential}
V(\psi) = \big[(\psi - B)^2 + A\big] e^{-\lambda \psi} ,
\end{equation}
characterized by the parameters $A$, $B$, and $\lambda$. An appealing feature of the AS potential is that it unifies early and late dark energy within a single scalar field, avoiding the need for two separate components. For example, the contribution at early times is modulated by the parameter $\lambda$, while $A$ and $B$ regulate the location and depth of the minimum that drives the onset of late-time acceleration~\cite{Adil:2022hkj, Garcia-Arroyo:2024tqq}.  
\\

\noindent
\textbf{Interaction Kernel.}\\
In order to solve the background and linear KG equations, it is necessary to specify the interaction kernel $Q$.
As an example,  in  \cite{Garcia-Arroyo:2024tqq} we considered a phenomenological kernel proportional to the  derivatives of each field
\begin{equation}
 Q = \beta \,\dot{\phi}\,\dot{\psi},
\end{equation}
which allows for a bidirectional exchange of energy density between the two scalar fields.
However, this interaction  is efficient only while both fields have non-negligible kinetic energy; that is, once the average contribution becomes small, the fields evolve approximately as decoupled components, which is indeed the behavior found on average.  As a consequence, the effects of the interaction become barely noticeable, and current observations are not significantly sensitive to the coupling parameter $\beta$. This motivates the exploration of alternative interaction kernels.
\\

In fluid descriptions of interacting DM--DE models, the interaction is commonly written as the product of a characteristic interaction rate, typically proportional to the Hubble expansion rate $H(z)$ or its value today $H_0$, and a function of dark-sector energy densities~\cite{ Valiviita:2008iv, Wang:2016lxa, DiValentino:2019ffd, Carrion:2024itc, Giare:2024smz}. Representative examples include linear kernels,
\begin{equation}\label{eq:linear_int}
    Q = 3H\gamma\rho_i, \qquad Q = 3H_0\gamma\rho_i, \qquad i={\rm dm,de},
\end{equation}
as well as nonlinear extensions~\cite{Li:2018jiu,Wang:2024vmw},

\begin{subequations}\label{eq:nonlinear-kernels}
\begin{align}
Q &= 3H\gamma\,
    \frac{\rho_{\rm dm}^{\,\alpha}\rho_{\rm de}^{\,\beta}}
         {\rho_{\rm crit}^{\,\sigma}},
&
Q &= 3H_0\gamma\,
    \frac{\rho_{\rm dm}^{\,\alpha}\rho_{\rm de}^{\,\beta}}
         {\rho_{\rm crit}^{\,\sigma}},
\label{eq:nonlinear-kernels-a}
\\
Q &= 3H\gamma\,
    \frac{\rho_{\rm dm}^{\,\alpha}\rho_{\rm de}^{\,\beta}}
         {(\rho_{\rm dm}+\rho_{\rm de})^\sigma},
&
Q &= 3H_0\gamma\,
    \frac{\rho_{\rm dm}^{\,\alpha}\rho_{\rm de}^{\,\beta}}
         {(\rho_{\rm dm}+\rho_{\rm de})^\sigma},
\label{eq:nonlinear-kernels-b}
\end{align}
\end{subequations}
where $\rho_{\rm crit}$ is the critical density, $\gamma$ the coupling parameter, and the exponents ($\alpha$, $\beta$, $\sigma$) define a broad family of interaction kernels, while different choices for the denominator provide alternative normalization schemes in which other energy scales can be considered. Alternative approaches that avoid explicit parametric specification of the kernel function, instead reconstruct the kernel directly from the data using non-parametric methods~\cite{Escamilla:2023shf}.

When one or both dark components are represented by scalar fields, the kernel is often built from field velocities. In coupled-quintessence models, for instance, interactions proportional to $\dot{\psi}\rho_{\rm dm}$ are commonly considered~\cite{Amendola:1999er, Xia:2013nua}, while decaying scalar field models introduce a decaying rate through terms that are proportional to the kinetic energy~\cite{Boehmer:2008av}  or kernels similar to the fluid DE models ~\cite{Pourtsidou:2013nha, Costa:2014pba, Boehmer:2015kta, Boehmer:2015sha, Perez:2021cvg, vandeBruck:2022xbk, Roy:2023uhc, Aboubrahim:2024spa, Kritpetch:2024rgi}, just by replacing the densities in Eqs.~\ref{eq:linear_int}--\ref{eq:nonlinear-kernels} with those of the scalar fields $(\rho_\phi,\, \rho_\psi)$. More generally, scalar field interactions can depend on the field variables, including their potentials, and may therefore become relevant in different dynamical regimes.
\\

Here, we generalize the nonlinear kernels, Eqs.~\ref{eq:nonlinear-kernels-a}-\ref{eq:nonlinear-kernels-b}, by replacing the density dependence in the numerator with the combination ($\rho\pm p$). For the scalar fields considered, $\rho_\phi+p_\phi=\dot{\phi}^{2}$  $(\rho_\psi+p_\psi=\dot{\psi}^{2})$, while  $\rho_\phi- p_\phi= 2 V(\phi)$  $(\rho_\psi- p_\psi= 2 V(\psi))$, such that the corresponding sign isolates the kinetic and potential contributions, respectively. In this work, we consider the positive-sign combination that naturally appears in the continuity equations and characterizes the dynamical evolution of each component. We further allow for different choices of the  normalization scale and Hubble prefactor, leading to a broad family of interaction kernels.

\begin{figure}
    \centering
    \includegraphics[width=0.9\linewidth]{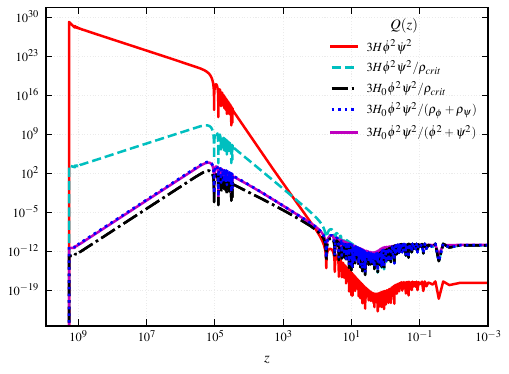}
    \caption{Evolution of the interaction kernel, $Q(z)$, for different choices of the Hubble prefactor and normalization scale. Line styles and colors identify the interaction kernels, as indicated by the labels.}
    \label{fig:kernel_comparison}
\end{figure}

Fig.~\ref{fig:kernel_comparison} compares several representative implementations of the interaction kernel, $Q(z)$, all evaluated with $\gamma=1$. First, the red curve, corresponding to an interaction proportional to $H(z)$ without a normalizing scale, shows that the interaction becomes excessively large at early times and subsequently decays much faster than the normalized cases, following the evolution of $H$. Next, by introducing the normalization scale $\rho_{\rm crit} \propto H^2$ (cyan curve), this abrupt early-time behavior is suppressed; however, it is still too high to be phenomenologically viable. We then replace $H(z)$ with $H_0$ while keeping  $\rho_{\rm crit}$ in the denominator (black curve). This yields a more natural evolution: the interaction starts from negligible values in the early Universe, $H(z)\gg H_0$, then it grows to a moderate maximum as the kinetic energies of both scalar fields become simultaneously significant, and subsequently decreases towards a small oscillatory regime as the dark-energy field approaches $w_\psi\simeq-1$. This consideration motivates the adoption of interaction kernels proportional to $H_0$.
 \\

Finally, to assess the impact of the normalization scale, the blue and magenta curves, corresponding to the choices of $\rho_{\phi}+\rho_{\psi}$ and $\dot \phi ^2 + \dot \psi ^2$, respectively, are nearly indistinguishable from each other and remain comparable in shape and magnitude to the case $\rho_{\rm crit}$, although they exhibit a slightly steeper decay after reaching their maximum. Therefore, the interaction is only weakly sensitive to the choice of normalization. 
Consequently, hereafter we adopt $\rho_{\rm crit}$ as the normalization scale and $H_0$ as the Hubble prefactor.
The resulting generalized interaction kernel is 
\begin{equation}
   Q = 3H_0\gamma\frac{(\rho_{\rm dm}+p_{\rm dm})^{\alpha}(\rho_{\rm de}+p_{\rm de})^{\beta}}{\rho_{\rm crit}^{\, \sigma}} = 3H_0\gamma\,\frac{\dot{\phi}^{2\alpha}\dot{\psi}^{2\beta}}{\rho_{\rm crit}^{\, \sigma}} \, .
\end{equation}
 Since  $3H_0\gamma$ is a constant, we define $\Gamma=3H_0\gamma$ as the interaction parameter, with dimensions of $H_0$. 
 For simplicity, we adopt the particular choice of $\alpha=\beta=\sigma=1$, resulting in
\begin{equation}\label{eq:kernel_quad}
    Q= \Gamma \frac{\dot\phi^2 \dot\psi^2}{\rho_{\rm crit}}\, .
\end{equation}
 This form of the interaction can also be obtained by considering that each scalar field decays at a certain rate; imposing conservation of the total energy of the dark-sector then leads to an interaction of this type~\cite{Garcia-Arroyo:2024tqq}. 
In contrast to the linear interaction, whose average contribution tends to vanish once the fields enter the rapid oscillation regime, the quadratic form remains non-zero on average, allowing the interaction to produce sustained effects during the cosmological evolution.

\begin{figure}
    \centering
    \includegraphics[width=0.9\linewidth]{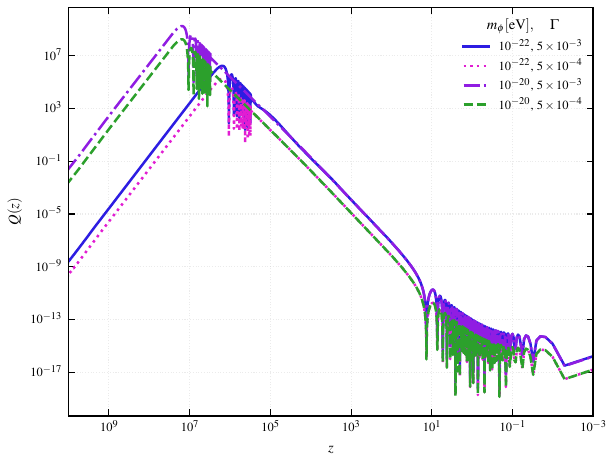}
    \caption{Evolution of the interaction kernel, Eq.~\ref{eq:kernel_quad}, for different $m_\phi$ and $\Gamma$. All remaining parameters are fixed as listed in Column~III of Table~\ref{tab:results}, with $\lambda=35$ and $A\lambda^2=0.26$. Different values are identified by the colored labels.}
    \label{fig:kernel_quad}
\end{figure}

The evolution of the considered kernel, Eq.~\ref{eq:kernel_quad}, depends on the dynamics of both the scalar fields and the parameter $\Gamma$.  Fig.~\ref{fig:kernel_quad} illustrates this dependence for two values of $m_\phi$ and two values of $\Gamma$, while keeping all other parameters fixed to the values reported in Column~III of Table~\ref{tab:results}. 
For a fixed mass, but varying $\Gamma$, the overall interaction is simply rescaled, whereas the SFDM mass determines both the amplitude of the peak and the onset of the first oscillatory regime, with larger masses producing higher peaks and shifting both features to earlier times. After this initial oscillatory stage, the dependence on the SFDM mass becomes negligible, and models with the same $\Gamma$ evolve almost identically.

\begin{figure*}[t]
      \includegraphics[width=0.32\linewidth]{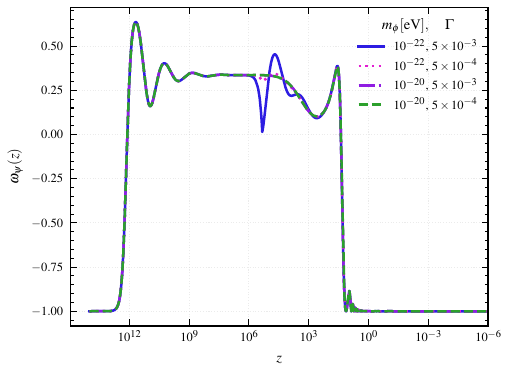}
      \includegraphics[width=0.32\linewidth]{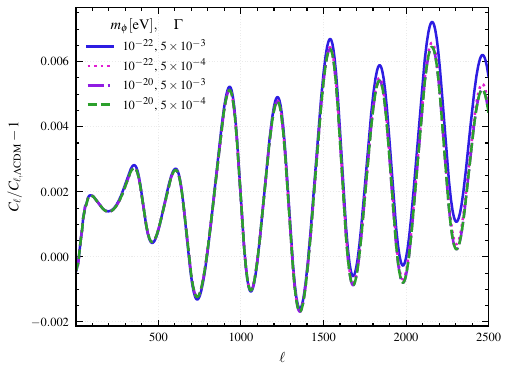} \includegraphics[width=0.32\linewidth]{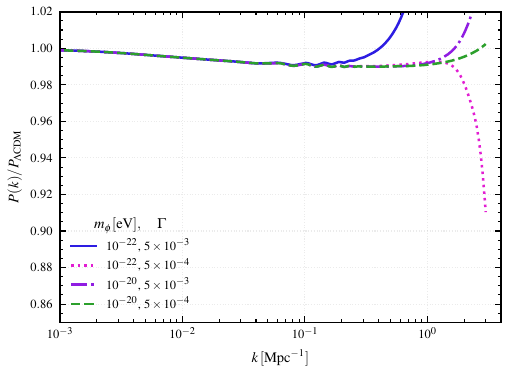}
\caption{Impact of the SFDM mass $m_\phi$ and the coupling $\Gamma$ parameters on cosmological observables. All remaining cosmological and model parameters are fixed to the values in Column~III of Table~\ref{tab:results}, with $\lambda = 35$ and $A\lambda^2 = 0.26$. \textit{Left}: Dark energy equation-of-state parameter. \textit{Middle}: Residual CMB temperature power spectrum with respect to $\Lambda$CDM. \textit{Right}: Ratio of the matter power spectrum  with respect to $\Lambda$CDM. Colors and line styles are as indicated by the labels.}
    \label{fig:three_panels}
\end{figure*}

Finally, Fig.~\ref{fig:three_panels} summarizes the cosmological signatures of the quadratic coupling  through the dark energy equation of state ($\omega_{\psi}$), the residual CMB temperature power spectrum, and the matter power spectrum ratio with respect to $\Lambda$CDM, using the same parameter combinations as in Fig.~\ref{fig:kernel_quad}.  %

The left panel shows the evolution of $\omega_{\psi}$ and illustrates how the coupling parameter $\Gamma$ induces oscillations around the standard EoS behavior once the SFDM field leaves its frozen regime and starts oscillating. These features become more pronounced for lighter SFDM masses and larger values of $|\Gamma|$ and are only noticeable in the early Universe, whereas at late times, all models asymptotically converge to the AS behavior. 
The middle and right panels display the residual CMB temperature power spectrum and the ratio of the matter power spectrum (MPS), respectively. For the CMB, all parameter selections lead to comparable deviations from $\Lambda$CDM. However, for the lightest SFDM mass, $m_{\phi}\sim10^{-22}\mathrm{eV}$, and the largest coupling $\Gamma=5\times10^{-3}$, the deviations become significantly more pronounced, showing the sensitivity of the CMB to the model parameters.  

In contrast, the MPS ratio with respect to $\Lambda$CDM exhibits a clear dependence on both the mass of SFDM and the strength of the interaction. 
For a fixed SFDM mass, increasing $\Gamma$ enhances the matter power spectrum on small scales, whereas lighter SFDM masses shift this enhancement toward lower values of $k$. 
This behavior arises because the positive interaction transfers energy from dark energy to dark matter, enhancing structure growth on small scales, whereas a negative interaction transfers energy in the opposite direction, suppressing structure growth.
\\

The two–field framework developed here provides a  flexible description of interacting dark sector models, particularly when both components are treated as scalar fields. In this work, we adopt a quadratic potential for the SFDM and the Albrecht–Skordis potential for the DE field; nevertheless, the formalism and the code can accommodate alternative scalar field potentials together with the generalized family of interaction kernels, encompassing both linear and nonlinear interactions as particular cases.
The background and linear equations have been implemented in a modified version of \texttt{CLASS}, following the strategy established in~\cite{Garcia-Arroyo:2024tqq}. The code is publicly available\footnote{\url{https://github.com/gabygga/Interacting}}, and its framework can be straightforwardly modified to accommodate various dark sector models that are able to interact.

\section{Methodology}\label{sec:methodology}

To explore the parameter space, we performed a Markov Chain Monte Carlo (MCMC) analysis using a modified version of the publicly available \texttt{MontePython} code \cite{Brinckmann:2018cvx}. The parameter setup and datasets used in the analysis are briefly described in this section.

\subsection{Parameter Setup}
The cosmological and model parameters sampled during our analysis, along with their prior ranges, are summarized in Table \ref{tab:priors}. For instance, the baseline parameter set includes the present-day baryon density \(\omega_b\), the amplitude \(A_s\) and tilt \(n_s\) of the primordial power spectrum, the Hubble constant \(H_0\) and the optical depth to reionization \(\tau_{\rm reio}\). The angular size of the sound horizon \(\theta_s\) is treated as a derived parameter, and neutrinos are modeled with two effectively massless species and one massive state with \(m_\nu = 0.06\,\rm{eV}\), and \(N_{\rm eff} = 3.046\).

We extend this setup by considering three scenarios for the dark sector. First, we include the scalar-field DE component while retaining cold dark matter, denoted AS$(\psi)$+CDM. Second, we consider dark energy and dark matter described by scalar fields, AS$(\psi)$ + SFDM$(\phi)$, without direct coupling. Finally, we analyze the same two-field setup, including the direct coupling between the fields, controlled by Eq.~\ref{eq:kernel_quad}. 
These extensions introduce additional parameters associated with the scalar-field potentials. In the case of AS$(\psi)$, we vary the parameter $\lambda$ and the combination $A\lambda^2$ and, when present, the SFDM component is characterized by its current density $\Omega_{\phi}$ and its mass, parameterized as $\log_{10}(m_{\phi}/{\rm eV})$. In the interacting scenario, we sample over the coupling parameter $\Gamma$, expressed in  $\rm Mpc^{-1}$ units adopted internally by \texttt{CLASS}, hence the corresponding dimensionless coupling, $\gamma$, can be obtained straightforwardly as a derived parameter.

The choice of prior ranges for these scalar field parameters is motivated by both theoretical consistency and current observational bounds. For the SFDM sector, described by a quadratic potential, Eq. \ref{eq:qua_potential},  we adopt flat priors on  $\Omega_{\phi}\in [0.1, 0.7]$ and $\log_{10}(m_{\phi}/\rm{eV})\in[-25, -17]$, wide enough to play the role of dark matter while covering the fuzzy dark matter regime.  

For the DE component with the Albrecht--Skordis potential, Eq.~\ref{eq:AS_potential}, the existence of a late-time minimum requires the condition $1 - A\lambda^2 > 0$ \cite{Barrow:2000nc, Adil:2022hkj, Garcia-Arroyo:2024tqq}. Since this condition depends on the product $A\lambda^2$, fixing $A$ would also impose an upper bound on the slope parameter, that is, $\lambda < 1/\sqrt{A}$. For example, the commonly used value $A=0.0025$ restricts the sampling range to $\lambda < 20$. This restriction is not imposed by the data but by the parametrization, and can artificially limit the region in which the AS field approaches the $\Lambda$CDM-like regime. We therefore take $A\lambda^2$ as the sampled parameter, with a flat prior in the interval $[0.10,0.99]$, and vary $\lambda \in [8,75]$ independently. This parametrization enforces the existence of the minimum while allowing the data to determine how steep the potential can be. It also keeps the early dark energy fraction, approximately $\Omega_{\psi}^{\rm early}\sim 4/\lambda^2$, at the $\mathcal{O}(10^{-1})$ level or below, consistent with observational bounds \cite{Agrawal:2019lmo,CosmoVerseNetwork:2025alb}. The results obtained with fixed $A=0.0025$ are therefore used only as a reference case to connect with previous AS analyzes and to illustrate the impact of the parametrization; they are not adopted as our baseline constraints.
The parameter $B$ is treated as a derived parameter, determined by the requirement that at late times the scalar field $\psi$ approaches the minimum of the Albrecht--Skordis potential, such that $V_{\rm min}\simeq \rho_{\psi,0}=3M_{\rm Pl}^2H_0^2\Omega_\psi$, which isolating $B$ yields
\begin{displaymath}{
B = 
\frac{\sqrt{1-A\lambda^{2}}-1}{\lambda} \\
+\frac{1}{\lambda}
\ln\!\left[
\frac{\left(1-\sqrt{1-A\lambda^{2}}\right)^{2}+A\lambda^{2}}{3\lambda^{2}M_{\rm Pl}^{2}H_{0}^{2}\Omega_{\psi}}
\right].}
\end{displaymath}
Meanwhile, the interaction between the two fields is described by the coupling constant $\Gamma$, for which we adopt a broad non-informative prior.

\begin{table}[t]
\centering
\caption{Priors on the cosmological and model parameters used throughout the analysis.}
\begin{tabular}{lc}
\toprule
Parameter & Prior \\
\hline
$\omega_b$            & $\mathcal{U}[0.005,\,0.1]$ \\
$\ln(10^{10}A_s)$     & $\mathcal{U}[2.7,\,4.0]$ \\
$n_s$                 & $\mathcal{U}[0.9,\,1.1]$ \\
$H_0$ [km/s/Mpc]      & $\mathcal{U}[ 40,\, 120]$\\
$\tau_{\mathrm{reio}}$ & $\mathcal{U}[0.01,\,0.8]$ \\
\multicolumn{2}{c}{\dotfill}\\
$\Omega_{\mathrm{\phi}}$ & $\mathcal{U}[0.1,\,0.7]$ \\
$\log m_\phi\,[\mathrm{eV}]$ & $\mathcal{U}[{-25},\,{-17}]$ \\
$A\lambda^2$             & $\mathcal{U}[0.1,\,0.99]$ \\
$\lambda$             & $\mathcal{U}[8,\,75]$ \\
$\Gamma$               & $\mathcal{U}[-2\times10^{-3},\,4\times10^{-3}]$ \\
\hline
\end{tabular}
\label{tab:priors}
\end{table}

\subsection{Datasets}
We perform a Markov chain Monte Carlo (MCMC) analysis using a modified version of \texttt{CLASS} and \texttt{MontePython}, in combination with the following datasets:

\begin{itemize}
    \item \noindent
\textbf{Planck 2018}: hereafter simply referred to as \textit{Planck}; this dataset comprises the combined likelihoods of the temperature (TT), polarization (EE), and temperature--polarization cross-correlation (TE) power spectra, together with CMB lensing. Specifically, we employ the \texttt{Commander} and \texttt{SimAll} likelihoods for the low-$\ell$ temperature and polarization spectra (TTEE), the \texttt{Plik} likelihood for the high-$\ell$ auto and cross spectra (TTTEEE), and the \texttt{smica} lensing likelihood~\cite{Planck:2019nip}.

    \item \textbf{DESI BAO DR2}: hereafter simply referred to as DESI~\cite{DESI:2025zgx}; we use the BAO measurements from galaxies, quasars, and the Ly$\alpha$ forest at different effective redshifts, which provide several independent constraints on distance ratios and scaling parameters.
    
    \item \textbf{eBOSS DR14 \(\mathrm{Ly}\alpha\) forest}: 
    by simplicity referred to as \(\mathrm{Ly}\alpha\), this dataset is based on measurements of the one-dimensional flux power spectrum from quasar absorption lines in the eBOSS DR14 survey. 
    Rather than directly fitting the observed  \(\mathrm{Ly}\alpha\) flux power spectrum; we use the reconstructed linear matter power spectrum data at $z=0$ provided by~\cite{Chabanier:2019eai}.
    \item \textbf{Pantheon\(+\)SH0ES}: we use the compilation of Type Ia supernova data, comprising 1,701 light curves from 1,550 distinct SNe Ia. This dataset covers the redshift range $0.01 < z < 2.26$ and provides measurements of the distance modulus, with the associated covariance matrix that includes both statistical and systematic uncertainties \cite{Scolnic:2021amr, Brout:2022vxf}.
\end{itemize}

Throughout the analysis, we consider two combinations of the above data sets:

\begin{itemize}
    \item \textbf{Base} (Planck + DESI + \(\mathrm{Ly}\alpha\)):  Combines early universal constraints from Planck, the expansion history from DESI, and the linear matter power spectrum derived from $\mathrm{Ly}\alpha$ data, which provides sensitivity to the small--scale clustering where SFDM and EDE models are expected to exhibit their largest departures from  $\Lambda$CDM.
    \item \textbf{Base + SN} (Base+Pantheon+SH0ES): provide precise constraints on the late-time expansion and the local value of $H_0$, allowing consistency tests between early- and late-universe measurements.
\end{itemize}

These two combinations allow us to disentangle the impact of early dark energy and scalar field dark matter on both the early- and late-time Universe. In particular, the Base + SN data combination is critical for assessing the potential role of these models in addressing the Hubble tension.
\section{Results}\label{sec:results}

\begin{figure}[t]
    \centering
    \includegraphics[width=0.5\textwidth]{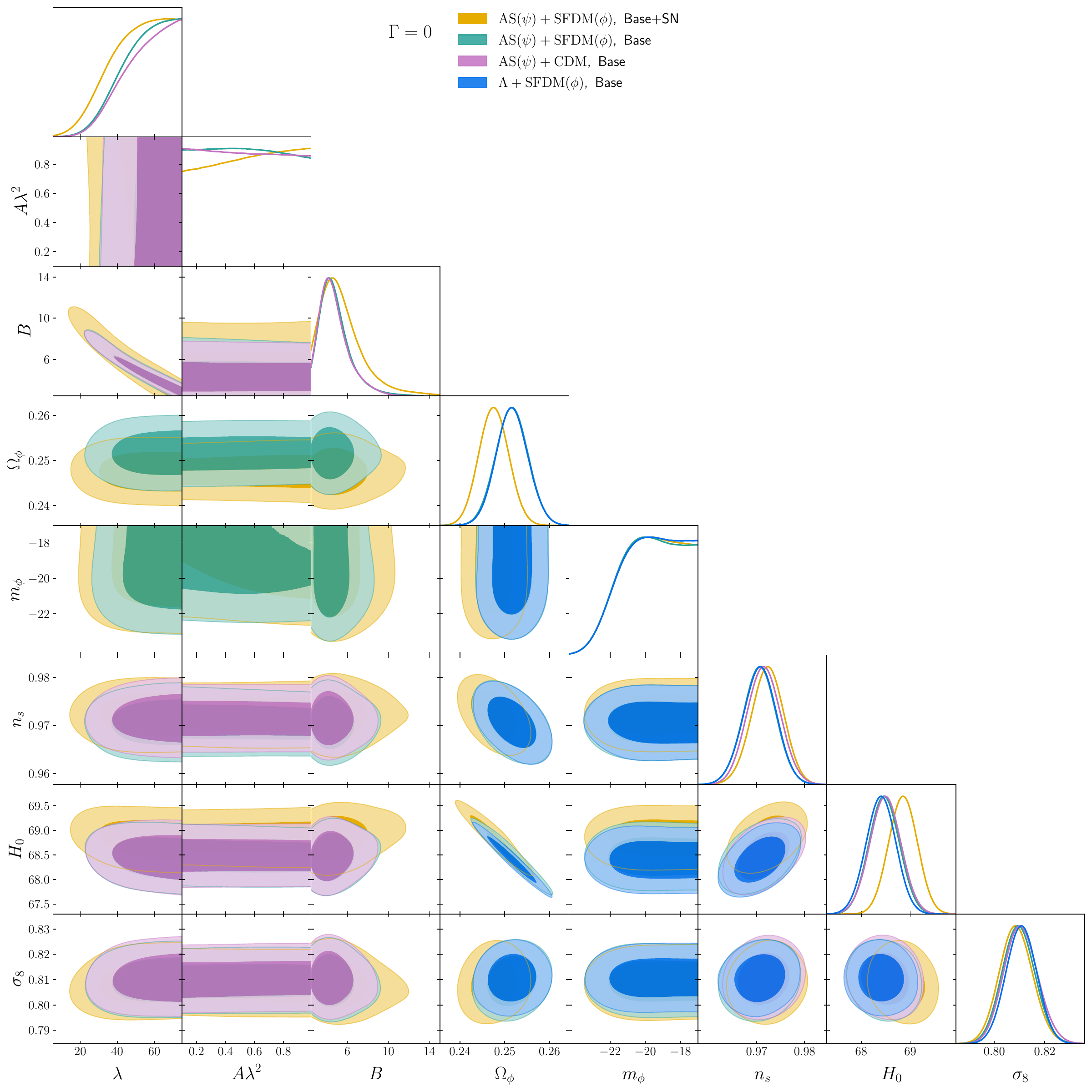}
    \caption{Posterior distributions of the most relevant cosmological and scalar field parameters at 68\% and 95\% c.l. for different scalar field scenarios; $\rm{AS}(\psi)$+CDM,  $\rm{AS}(\psi)$+$\rm{SFDM}(\phi)$ and $\Lambda$+$\rm{SFDM}(\phi)$, all without direct coupling. For the $\rm{AS}(\psi)$+CDM and $\Lambda$+$\rm{SFDM}(\phi)$ cases, only the Base dataset is shown for clarity, as the corresponding constraints are nearly indistinguishable from those obtained for the $\rm{AS}(\psi)$+$\rm{SFDM}(\phi)$ scenario. }
    \label{fig:triangle_nobeta}
\end{figure}

\begin{figure}[t]
    \centering
    \includegraphics[width=0.5\textwidth]{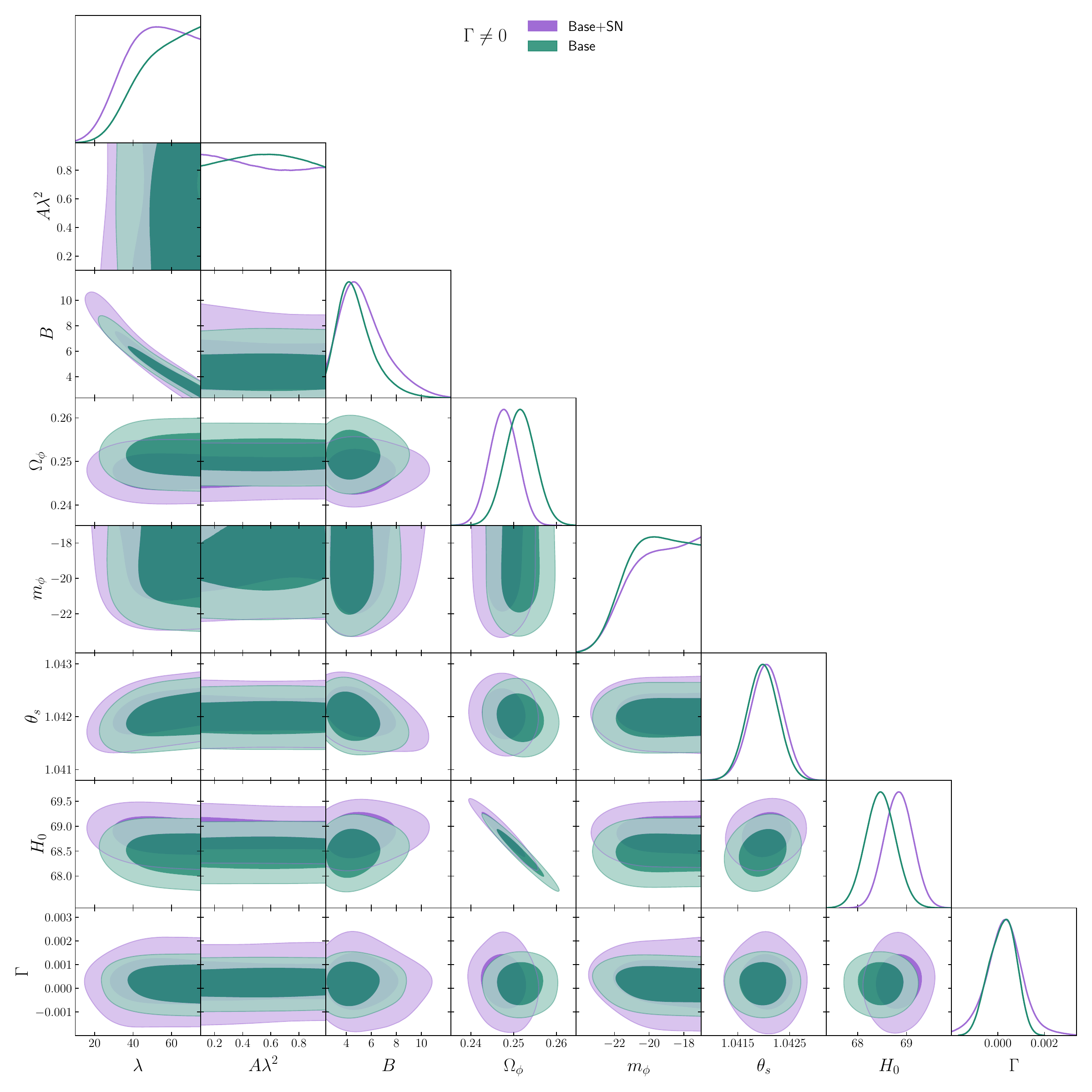}
    \caption{
Posterior distributions for the interacting $\rm{AS}(\psi)$+$\rm{SFDM}(\phi)$ model, including the standard cosmological parameters, the scalar-field parameters, and the coupling parameter $\Gamma$. Contours show the 68\% and 95\% c.l. for the Base and Base+SN dataset combinations.
}
\label{fig:triangle_beta}
\end{figure}

%
In this section, we present the main results of the  Bayesian inference performed for different combinations of data sets and theoretical scenarios. In Figs.~\ref{fig:triangle_nobeta} and~\ref{fig:triangle_beta}, we show the posterior distributions and 2D-contours at 68\% and 95\% confidence levels (c.l) for the most relevant parameters of the models, while in Tab.~\ref{tab:results} we present the mean parameter values with uncertainties reported at 68\% c.l. and lower and upper bounds at 95\% c.l.
For comparison, the corresponding $\Lambda\rm CDM$ constraints are also included.

\begin{table*}[t]
\centering
\footnotesize
\begin{tabular}{lccccccc}
\hline\hline
 {}&\multicolumn{2}{c}{$\Lambda\rm{CDM}$}&\multicolumn{1}{c}{$\rm AS(\psi)+\rm{CDM}$}&\multicolumn{2}{c}{$\rm AS(\psi)+SFDM(\phi)$}& \multicolumn{2}{c}{$\rm AS(\psi)+SFDM(\phi)+Q$}\\
Parameter &  Base & Base+SN&Base & Base & Base+SN& Base & Base+SN \\
\hline
$10^{-2}\omega{}_{b }$ &$2.253\pm 0.013$ & $2.264\pm 0.013$&$2.256\pm 0.013$ &$2.254\pm 0.012$ &$2.265\pm 0.012$ &$2.254\pm 0.013$ & $2.264\pm 0.013$\\
$\theta_s$ & $1.0421\pm 0.0003$& $1.0422\pm 0.0003$ & $1.0420\pm 0.0003$       & $1.0420\pm 0.0003 $ &$1.0421\pm 0.0003$ &$1.0420\pm 0.0003$& $1.0421\pm 0.0003$ \\
$\ln(10^{10}A_{s})$ & $3.059\pm 0.015$ & $3.063 \pm 0.016$ &$3.060\pm 0.015$ & $3.060\pm 0.015$ & $3.063\pm 0.015$ &$3.059\pm 0.013$ &$3.063\pm 0.015$\\
$n_s$  & $0.971 \pm 0.003$ & $0.972 \pm 0.003$& $0.972\pm 0.003$           & $0.971\pm 0.003$ & $0.972\pm 0.003$  &$ 0.971\pm 0.003$&$0.972\pm 0.003$ \\
$H_0$ [km/s/Mpc] & $68.41\pm 0.29$&$68.79\pm 0.28$ &$68.49\pm 0.31$  & $68.46\pm 0.30$ & $68.85\pm 0.28$ & $68.47\pm 0.30$ & $68.84\pm 0.28$\\
$\sigma_8$  & $0.810\pm 0.006$ & $0.810\pm 0.007$& $0.811\pm 0.006$      & $0.810\pm 0.006$ & $0.809\pm 0.006$ &$0.809\pm 0.006$ & $0.808\pm 0.006$ \\
\hline
$\Omega_c,\,\Omega_{\phi}$ &$0.252\pm0.003$ &$0.247 \pm 0.003$ &$0.252\pm 0.004$ & $0.252\pm 0.004$ & $0.248\pm 0.003$ &$0.252\pm 0.003$&$0.248\pm 0.003$\\
$\log_{10}(m_{\phi}/\rm{eV})$  &--- &--- &---        & $> -22.1$ & $> -22.2 $ &$> -22.1$ & $> -22.0$\\
$\lambda$ &--- &---&   $> 32.3$      & $> 32.1$ &$> 26.2$ & $> 32.2$ &  $> 27.4$\\
$A\lambda^2$ &--- &---& \rm{unconstrained} & \rm{unconstrained}& \rm{unconstrained} & \rm{unconstrained}& \rm{unconstrained} \\
$10^{-4}\Gamma$ &--- &---  & ---           &  --- & --- &$2.0^{+6.3}_{-5.4}$ & $2.5\pm 7.4$  \\
\multicolumn{8}{c}{\dotfill} \\
$\chi ^2_{\rm min}$ &$2806.94$ &$4128.36$ &$2807.4$ & $2807.5$ & $4128.2$& $2807.86$& $4128.7$\\
\hline\hline
\end{tabular}
\caption{
Posterior constraints on the cosmological and scalar-field parameters for the models considered in this work. Mean values and uncertainties are quoted at 68\% c.l., while lower bounds are reported at 95\% c.l. Columns I and II display the constraints for the $\Lambda$CDM model. The third column corresponds to the $\rm AS(\psi)+CDM$ model using the Base dataset. The next two columns show the $\rm AS(\psi)+SFDM(\phi)$ case without direct coupling for the Base and Base+SN datasets, and the last two columns show the interacting $\rm AS(\psi)+SFDM(\phi)+Q$ case for the same dataset combinations. Here, $Q$ denotes the coupling kernel between the scalar fields, while $\Gamma$ is the sampled coupling parameter. The last row report the minimum $\chi^2$ values for each model and dataset.}
\label{tab:results}
\end{table*}

To assess the impact of introducing a scalar field in either the dark matter or the dark energy sector, we first compare the three uncoupled configurations shown in Fig.~\ref{fig:triangle_nobeta}: $\rm{AS}(\psi)$+CDM, $\rm{AS}(\psi)$+$\rm{SFDM}(\phi)$, and $\Lambda$+$\rm{SFDM}(\phi)$. We begin by replacing the standard CDM fluid with a scalar-field dark matter component while keeping the AS field as the dark energy sector. The resulting constraints are nearly identical in both cases. In particular, the baseline cosmological parameters, the lower bound on $\lambda$, and the minimum $\chi^2$ values reported in Tab.~\ref{tab:results} show no significant dependence on whether the dark matter sector is described by CDM or SFDM. Moreover, in the scenario of two scalar fields, the posterior contours do not reveal any relevant degeneracy between the SFDM and AS parameters. This suggests that the additional scalar degree of freedom in the dark matter sector does not significantly modify the dynamics of the AS field beyond its gravitational contribution, and vise versa, in agreement with our previous findings~\cite{Garcia-Arroyo:2024tqq}. Similarly, replacing the AS field with a cosmological constant while keeping the SFDM component leads to virtually identical constraints on the SFDM parameters. We therefore do not further discuss this case.

\begin{figure}[t]
    \centering
    \includegraphics[width=0.4\textwidth]{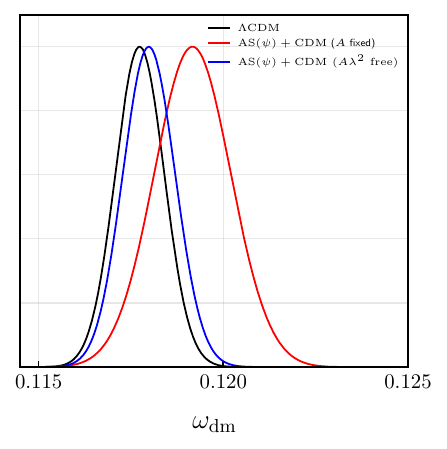}
    \caption{
    Posterior distributions of $\omega_{\rm dm}$ for $\Lambda$CDM and $\rm{AS}(\psi)$+CDM using the Base dataset. The fixed-$A$ curve, with $A=0.0025$ (red), is shown only as a reference case: it effectively imposes $\lambda < 20$. Our adopted parametrization samples $A\lambda^2$ instead (blue), allowing the data to prefer larger values of $\lambda$ and yielding an $\omega_{\rm dm}$ posterior closer to $\Lambda$CDM. 
    }
    \label{fig:omega_dm}
\end{figure}

For these kinds of models, it is useful to examine the physical abundance of dark matter $\omega_{\rm dm}$, since EDE models often compensate for the extra early-time DE density with a larger DM contribution to preserve the acoustic scale. Here $\omega_{\rm dm}=(H_0/100)^2\Omega_c$ for CDM and $\omega_{\rm dm}=(H_0/100)^2\Omega_{\phi}$ for SFDM. For the Base dataset, Tab.~\ref{tab:results} gives $\Omega_c\simeq\Omega_{\phi}\simeq0.252$ and $H_0\simeq68.5\,\rm{km\,s^{-1}\,Mpc^{-1}}$ in both configurations, leading to very similar $\omega_{\rm dm}$ posteriors that remain close to the corresponding $\Lambda$CDM result, as shown in Fig.~\ref{fig:omega_dm}. The same figure also illustrates the role of the AS parametrization by comparing our adopted sampling of $A\lambda^2$ with a fixed-$A$ reference case. This comparison shows that a restricted AS parametrization can shift the inferred matter abundance: fixing $A=0.0025$ imposes $\lambda<20$ and leads to a larger early DE contribution, while sampling $A\lambda^2$ allows the data to prefer larger values of $\lambda$, reducing $\Omega_{\psi}^{\rm early}\sim4/\lambda^2$ and moving $\omega_{\rm dm}$ closer to its $\Lambda$CDM value.

Next, we focus on the scenario $\rm{AS}(\psi)$ + $\rm{SFDM}(\phi)$ and compare the combinations of Base and Base+SN datasets. Since supernovae mainly constrain the late-time expansion history, they are expected to be most sensitive to the low-redshift behavior of the AS field, rather than its early-time contribution. As shown in Fig.~\ref{fig:triangle_nobeta} and in columns (2) and (3) of Tab.~\ref{tab:results}, adding SNeIa data mildly relaxes the lower bound on the AS slope, from $\lambda>32.1$ for Base to $\lambda>26.2$ for Base+SN. Interestingly, for both datasets the preferred region remains above $\lambda=20$, the maximum value allowed by the fixed choice $A=0.0025$ discussed above. Because higher values of $\lambda$ correspond to a smaller early dark energy fraction and to a behavior closer to $\Lambda$CDM, relaxation in the Base+SN case indirectly allows for a slightly larger early contribution, increasing from $\Omega_{\psi}^{\rm early}\lesssim0.0039$ to $\Omega_{\psi}^{\rm early}\lesssim0.0058$. In contrast, the SFDM sector remains essentially stable: the lower bound on $m_\phi$ and the inferred value of $\Omega_\phi$ change only slightly between the two dataset combinations. The remaining standard cosmological parameters are also weakly affected, the most visible shift being the expected late-time adjustment in $H_0$.
\\

\begin{figure}
    \centering
    \includegraphics[width=0.4\textwidth]{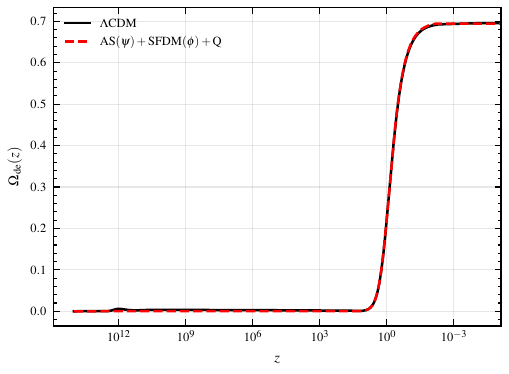}
    \caption{Evolution of the dark-energy density parameter, $\Omega_{\rm de}$, for the best-fit $\Lambda$CDM and AS+SFDM+$Q$ cosmologies.}
    \label{fig:omega_de_comparison}
\end{figure}

We now turn to the most general case, in which the two scalar fields are allowed to exchange energy directly through the interaction parameter $\Gamma$. As shown in Fig.~\ref{fig:triangle_beta} and in Tab.~\ref{tab:results}, introducing this coupling leaves the main cosmological constraints essentially unchanged with respect to the uncoupled case. Nevertheless, the posterior peak of $\Gamma$ is not located at the zero position; instead, it exhibits a mild preference for positive values for both dataset combinations, with mean values of the order $10^{-4}$, corresponding to a net energy transfer from the AS field to the SFDM sector. However, the 95\% c.l. region also includes zero and negative values of $\Gamma$, indicating that the opposite direction of energy flow is still allowed.  
It is worth noting that the preferred values of $\Gamma$ keep the $\rm{AS}(\psi)$  field close to its radiation-tracking solution. As illustrated by the EoS evolution in the left panel of Fig~\ref{fig:three_panels}, this behavior is progressively disturbed as $\Gamma$ increases. Despite these modifications to the field dynamics, their impact on the background evolution remains small for the mean $\rm AS(\psi)+SFDM(\phi)+Q$ parameter values. This is further illustrated in Fig.~\ref{fig:omega_de_comparison}, where the evolution of $\Omega_{\rm de}$ for the best-fit AS+SFDM+$Q$ cosmology closely follows that of the best-fit $\Lambda$CDM model.

Given that no statistically significant correlations were identified among the AS($\psi$), SFDM($\phi$), and coupling sectors, a common tendency emerges across all scenarios considered: the scalar-field parameters are driven toward their $\Lambda$CDM-like limits. For example, the SFDM mass is bounded to be sufficiently large $\log_{10}(m_\phi/\rm{eV})>-22$\cite{Urena-Lopez:2023ngt}, while the AS slope remains large as well $\lambda\gtrsim26$, which implies that the model yields only a negligible contribution from early dark energy.

Regarding correlations with the baseline cosmological parameters, the dominant trends are the expected ones. In particular, $H_0$ is anticorrelated with the dark matter abundance, which in the SFDM scenarios is encoded in $\Omega_\phi$. Thus, when SNeIa are included, the slight upward shift in the inferred value of $H_0$ is accompanied by a corresponding decrease in $\Omega_\phi$. However, even with this shift, $H_0$ remains close to the Planck-$\Lambda$CDM range, reaching only $H_0\simeq68.9\,\rm{km\,s^{-1}\,Mpc^{-1}}$, and therefore does not produce the upward displacement usually required to alleviate the tension on $H_0$~\cite{Pettorino:2013ia, Planck:2019nip, Riess:2021jrx}. Similarly, the parameter $\sigma_8$ remains approximately $0.81$ across all scenarios considered; therefore, the most general model does not increase the tension associated with the amplitude of matter fluctuations. Finally, the minimum values of $\chi^2$ remain very close to those of the corresponding $\Lambda$CDM fits for the same datasets.

\section{Conclusions}\label{sec:conclusions}

This work presents a unified numerical framework for modeling interacting scalar fields within the dark sector. The framework builds upon the pre-existing SFDM implementation ~\cite{Cedeno:2017sou} together with the quintessence module in \texttt{CLASS},  adapting and extending both implementations to consistently describe interacting scalar-field cosmologies through a broad family of interaction kernels. Although the present implementation focuses on two interacting scalar fields, the underlying formalism is sufficiently general to describe interactions between quintessence and CDM, SFDM and a dark-energy fluid, or, more generally, two interacting dark-sector fluids.

As a representative application, we applied this framework to constrain a model in which the dark matter component is described by SFDM with a quadratic potential and the dark-energy sector by the Albrecht--Skordis field, considering both the uncoupled case and a more general scenario in which the two scalar fields exchange energy through the nonlinear interaction kernel $Q\propto \Gamma (\rho_{\phi}+p_{\phi})(\rho_{\psi}+p_{\psi})$, which naturally switches off as the dark-energy component approaches the cosmological-constant regime $\omega_{\psi}\rightarrow -1$.

A common outcome is that, across all scenarios examined, the scalar-field parameters exhibit a systematic evolution toward the $\Lambda$CDM-like limiting values. The AS slope is constrained to exhibit large values, with $\lambda\gtrsim26$ even after including SNeIa data, implying only a small early dark energy contribution. At the same time, the SFDM mass is bounded to be sufficiently large, $\log_{10}(m_\phi/\rm{eV})>-22$, so that the dark matter sector remains close to the cold-dark-matter regime on the scales probed by the data. We also find no significant degeneracy between the AS and SFDM parameters, indicating that the two scalar sectors are constrained largely independently within the current combination of data sets.

The inclusion of SNeIa mainly affects the late-time sector, producing the expected mild increase in $H_0$ and a corresponding decrease in $\Omega_\phi$, consistent with the usual anticorrelation between the expansion rate and the abundance of dark matter. The posterior estimates of $H_0$ remain consistent with the Planck-$\Lambda$CDM constraints and thus do not considerably contribute to mitigating the $H_0$ tension. Likewise, the inferred value of $\sigma_8$ remains near $0.81$, indicating that the model stays consistent  with the observations of the amplitude of matter fluctuations.

In the interacting case, the posterior of $\Gamma$ does not peak at zero, but rather at positive values, which correspond to energy transfer from the AS field to the SFDM sector. Nevertheless, zero and negative values remain within the 95\% c.l. region, so the data do not select a unique direction for the energy flow. The coupling also does not lead to significant shifts in the baseline or scalar-field parameters. Finally, the minimum $\chi^2$ values remain close to those obtained in the corresponding $\Lambda$CDM fits, showing that the model is compatible with the data at the best-fit level, although the additional scalar-field and coupling parameters should be penalized in a full model-comparison analysis.
\\

As noted previously, the two-field framework offers a flexible and general parametrization of interacting dark-sector scenarios, particularly when both constituents are modeled as scalar fields. In the present analysis, we assumed a quadratic potential for the scalar-field dark matter (SFDM) component and the Albrecht–Skordis potential for the dark-energy (DE) field. Nonetheless, the underlying formalism and numerical implementation readily admit alternative choices of scalar-field potentials, as well as a generalized class of interaction kernels that include linear and nonlinear couplings as special cases. 
Planned future work will focus on extending the framework to incorporate a broader range of interacting dark-sector models and to assess their phenomenological viability.

\section*{Acknowledgments}
The authors acknowledge Ing. Francisco Bustos and Lic. Reyes Garc\'ia for their assistance with the High Performance Computing facilities at ICF-UNAM. G.G.A. acknowledges support from SECIHTI postdoctoral fellowship. The research of LAU-L was partially supported by Programa para el Desarrollo Profesional Docente; Dirección de Apoyo a la Investigación y al Posgrado, Universidad de Guanajuato; SECIHTI México under Grant No. CBF-2025-G-1327; and the Instituto Avanzado de Cosmología Collaboration. J.A.V. acknowledges support from UNAM-DGAPA-PAPIIT IN109126, IN110325, HTC project LANCAD-UNAM-DGTIC-477 and Cátedra de Investigación Marcos Moshinsky.

\bibliography{references}        

\end{document}